\documentstyle[12pt,psfig,aps]{revtex}
\begin{document}

\title{Characterization of a Low Frequency Power Spectral Density $f^{-\gamma}$
in a Threshold Model}
\author{Erika D'Ambrosio}
\address{LIGO Laboratory, Caltech M.C. 18-34 Pasadena CA91125}
\date{\today}

\maketitle
\begin{abstract}
This study investigates the modifications of the thermal spectrum, at
low frequency, induced by an external damping on a system in heat contact with
internal fluctuating impurities. Those impurities can move among locations and
their oscillations are associated with a loss function depending on the model.

The fluctuation properties of the system are provided by a potential function
shaped by wells, in such a way that jumps between the stationary positions are
allowed. The power spectral density associated with this dissipation mechanism
shows a $f^{-\gamma}$ tail.

The interest of this problem is that many systems are characterized by
a typical $f^{-\gamma}$ spectral tail at low frequency. The model presented in
this article is based on a threshold type behaviour and its generality allows 
applications in several fields.

The effects of an external force, introduced to produce damping,
are studied by using both analytical techniques and numerical simulations.

The results obtained with the present model show that
no reduction of the power spectral density is appreciable below the main peak
of the spectral density.\\
PACS Numbers: 05.40-a, 05.10-a
\end{abstract}
\thispagestyle{empty}
\pacs{05.40-a, 05.10-a}
\setcounter{page}{0}
\newpage
\section{Introduction}

Since gravitational wave interferometers will sense the passage of a 
gravitational wave as a difference in the light path between mirrors in the 
two arms, every kind of excitation is a noise source. Cooling the mirrors 
of these detectors, by measuring and controlling with a feedback loop the 
principal fundamental motions that may be thermally excited, may offer a 
means to reduce thermal noise. 

With this motivation in mind, we investigated models that generate frequency 
spectra with a tail $f^{-\gamma}$ for $f\rightarrow0$. This dependence is 
found in many completely different physical systems, 
suggesting a possible underlying simple mechanism.
The typical physical parameters are different for each of them, and especially
the characteristic microscopic quantities.
But many classes of phenomena are characterized by the same value of $\gamma$.

Indeed the responsive behaviour of every system is classified according to the
parameters of the phenomenological equations by which the main characteristics
of its evolution and interactions are described.
Moreover the contribution of the microscopic degrees of freedom is absorbed in
the definition of the parameters so that it is possible to describe the motion
of a system in terms of a phenomenologial equation for the macroscopic degrees
of freedom. The microscopic degrees of freedom motions are integrated over as
in averages and their effect in the dynamics of the system is contained in the
physical constants. For example in high energy physics the masses are corrected
for those particles which are not isolated but interacting with other ones, and
since measurements imply interactions, the measured values are the renormalized
values instead of the bare values. Similarly applications of those
principles may be found in low energy physics \cite{Ma,ZJ}.
A typical example is the up-conversion of resonant frequencies, related with a
non-linear perturbation term that affects both the bare value of the resonant
frequency and its width \cite{ED}.

In this paper a very simple model is studied whose basic characteristic is the
switching between equilibrium configurations across energy barriers. There are
no other requirements and this generality is the main feature of this flexible
model that can be easily adapted for a lot of different physical phenomena. 
The principal interest has been the investigation of the influence 
on the spectral function and particularly on the typical low frequency tail
of a damping force.

\section{The Model}

For those materials characterized by a quality factor that may be increased by
decreasing the amount of impurities, a reasonable picture is that the motion of
those degrees of freedom may generate a fluctuation-dissipation process across
the sites in the crystal. In other words the impurities are supposed to move 
following a random walk.

The energy of those impurities is related with their temperature; in cryogenic
conditions they hardly
can pass through the potential barrier and change their location. The
probability of a jump from one site to another depends upon $T$. The system is
characterized by its statistical properties and the temperature is defined as
a stationary thermodynamical quantity. 

\begin{figure}
\psfig{file=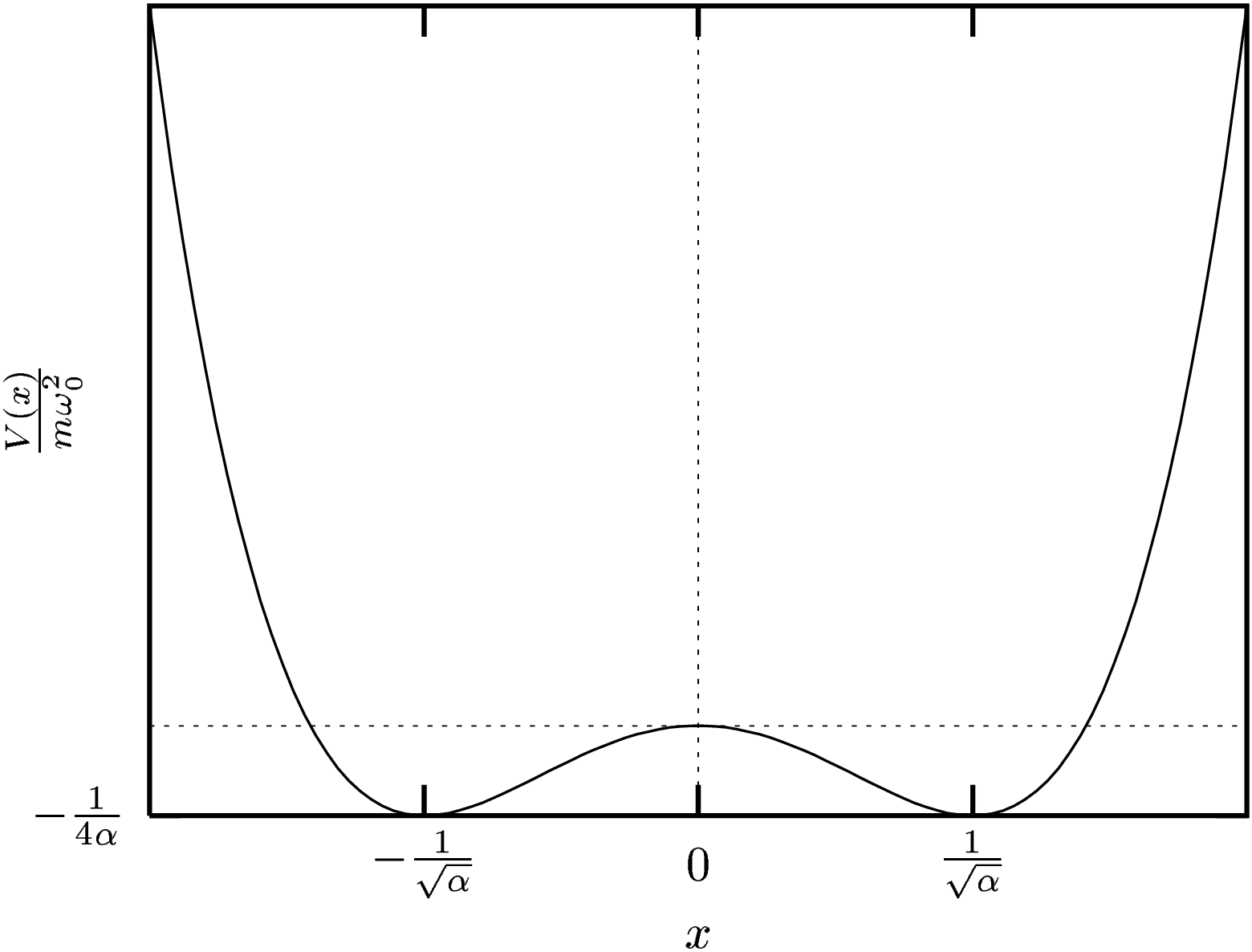,width=500pt,angle=-90}
\caption{The potential function $\sim-{x^2\over2}+\alpha{x^4\over4}$. 
The wells are farther apart if $\alpha$ decreases. Their depth depends on 
$\alpha$ as well, and scaling $x$ so that every quantity becomes dimensionless
gives $\alpha'\sim\alpha T$. Thus if $T$ increases the results are equivalent
to flattening the wells}
\label{fig:aa}
\end{figure}

A simple random walk model may
be considered to represent the switching between the equilibrium positions for
a two well potential function 
$$V(x)\sim{x^2\over2}(-1+{\alpha x^2\over2})\qquad$$
It has two minima for $x=\pm\sqrt{1\over\alpha}$ pointed out in 
Fig.\ref{fig:aa}.

Increasing the constant $\alpha$
implies a smaller distance between the two wells. From the Taylor expansion the
following expression is achieved
\begin{equation}
V(x)=m\omega_{0}^{2}[-{1\over{4\alpha}}+(x\pm{1\over\sqrt{\alpha}})^2+\ldots]
\end{equation}
for $x$ near the two minima. The constant $m\omega_{0}^{2}$ has been 
introduced so that $\sqrt{V''(x)\over m}$ has the physically right dimension 
of frequency.
In the phase space
\begin{eqnarray*}
\dot x & = & {p\over m} \\
\dot p & = & m\omega_{0}^{2}x(1-\alpha x^3)
\end{eqnarray*}
whose solutions represent the trajectories that depend on the initial conditions.

The problem may be treated perturbatively near the two minima. 

Using the linear
approximation for small oscillations a stochastic force may be introduced that
satisfies the fluctuation-dissipation theorem. 

In other words, the deterministic part of this force and the random component
are related in such a way that the exchanged energies are balanced.

The resulting equation is
$$m\ddot x+{m\omega_0\over Q}\dot x+2m\omega_{0}^{2}(x\pm{1\over\sqrt{\alpha}})
=\sqrt{2m\omega_0 K_BT\over Q}\xi(t)\qquad <\xi(t)\xi(t')>=\delta(t-t')$$
near the two minima $\mp\alpha^{-{1\over2}}$. 
The constant $Q$ is so large that underdamped conditions are always satisfied.

In order to take into account the ratio of the thermal equilibrium energy over
the local maxima in the potential function, a scale factor may be used in 
order to recover the equation
\begin{equation}
\ddot x'+{\omega_0\over Q}\dot x'+2\omega_{0}^{2}(x'\pm{1\over\sqrt{\alpha'}})
=\omega_0\sqrt{2\omega_0\over Q}\xi(t)   \label{eq:aa}
\end{equation}
where $x'=x\sqrt{m\omega_{0}^{2}\over{K_BT}}$ is dimensionless. 

As a consequence,
$\alpha$, that had the same dimensionality as $x^2$, has become dimensionless.

It
is now easy to see that flat wells and high temperature are equivalent and the
constant $\alpha'=\alpha{K_BT\over{m\omega_{0}^{2}}}$ sums up the two effects.

All the comments referred to Fig.\ref{fig:aa} are valid for $\alpha'$ as well.

More intutively the scaling of the x-axis by squeezing or stretching gives the
potential function a different shape corresponding to making the wells nearer 
or farther, deeper or flatter.

Following an expansion of the equation of motion near 
$\mp\alpha'^{-{1\over2}}$ 
\begin{eqnarray*}
x'(f)&=&G(f)[\omega_0\sqrt{2\omega_0\over Q}\xi(f)\mp 3\sqrt{\alpha'}
\omega_{0}^{2}\int_{-\infty}^{\infty}\mbox{d}f'x'(f')x'(f-f')+ \\ && 
\omega_{0}^{2}\alpha'\int_{-\infty}^{\infty}\mbox{d}f'
\int_{-\infty}^{\infty}\mbox{d}f''x'(f')x'(f'')x'(f-f'-f'')]
\end{eqnarray*}
where the origin of $x'$ has been fixed in $\pm{1\over\sqrt{\alpha'}}$. It may
be noted that $x'$ appears also on the right side of the equation and this 
means that this definition is implicit. Using a perturbative expansion gives
\begin{eqnarray*}
x'(f)&=&x'_0(f)+ G(f)[\omega_0\sqrt{2\omega_0\over Q}\xi(f)\mp 3\sqrt{\alpha'}
\omega_{0}^{2}\int_{-\infty}^{\infty}\mbox{d}f'x'_0(f')x'_0(f-f')+ \\ & &
\omega_{0}^{2}\alpha'\int_{-\infty}^{\infty}\mbox{d}f'\int_{-\infty}^{\infty}
\mbox{d}f''x'_0(f')x'_0(f'')x'_0(f-f'-f'')+ \\ & & 18\alpha\omega_{0}^{2}
\int_{-\infty}^{\infty}\mbox{d}f'x'_0(f-f')\omega_{0}^{2}G(f')
\int_{-\infty}^{\infty}\mbox{d}f''x'_0(f'')x'_0(f'-f'')+\ldots]
\end{eqnarray*}
where $x'_0$ represents the solution in the zero order approximation. 

If $x'_0$ is considered and the two point correlation function is calculated,
the associated integral over frequencies is
$$\int_{-\infty}^{\infty}\mbox{d}f\int_{-\infty}^{\infty}\mbox{d}f'<x'_0(f)
x'_0(f')>=\int_{-\infty}^{\infty}\mbox{d}f\int_{-\infty}^{\infty}\mbox{d}f'S(f)
\delta(f+f')={1\over2}$$
for each of the two linearly approximated solutions near the minima. 

The function $S(f)$ is called power spectral density. Taking into consideration higher order
terms for $x'$ gives corrections to $S(f)$. For the property $<\xi>=0$ the term
representing the first correction of $S(f)$ is proportional to $\alpha'$. It is
\begin{eqnarray*}
\lefteqn{<x'(f)x'(f')>-\delta(f+f')S(f)=} \\ &&
\delta(f+f')[9\alpha'\omega_0QS(f)
\int_{-\infty}^{\infty}\mbox{d}f_1S(f_1)S(f-f_1)+ \\ & & 
6\omega_{0}^{2}\alpha' S(f)\Re G(f)\int_{-\infty}^{\infty}\mbox{d}f_1S(f_1)+\\
&& 18\omega_{0}^{2}\alpha' S(f)\Re G(f)\int_{-\infty}^{\infty}\mbox{d}f_1S(f_1)
+\\ && 72\omega_{0}^{4}\alpha' S(f)\int_{-\infty}^{\infty}\mbox{d}f_1S(f_1)\Re
[G(f)G(f-f_1)]+\ldots]
\end{eqnarray*}
where $G(f)=(-\omega^2+{i\omega\omega_0\over Q}+2\omega_{0}^{2})^{-1}$. 

All the
terms may be easily derived using the Feynman diagram technique. In this paper
indeed the analogies between a wave equation and a heat equation are used and
the solution $x'(f)$ is represented and expanded in a graphic way. The analogy
with the Feynman diagram techniques is defined by associating a line with the
Green's function $G(f)$ and a cross with the force $\xi(f)$. When
two crosses are combined together a $\delta(f+f')$ arises. 
The iterative method of getting the terms of the expansion consists of 
substituting for the solution in the implicit definition at any order, 
the expression corresponding to the lower one in $\alpha'$. 

When $\alpha'=0$ the power spectral density may be calculated for each one of
the two linear approximations
$$S(f)=CG(f)G(-f)\qquad C={2\omega_{0}^{3}\over Q}\qquad.$$ 

Integrating over all frequencies $S(f)$ gives ${1\over2}$ for each of the two 
zero order expansions. This
is due to the fluctuation-dissipation theorem that links the dissipative force
to the constant in front of $\xi$ in the motion equation. 

\begin{figure}
\psfig{file=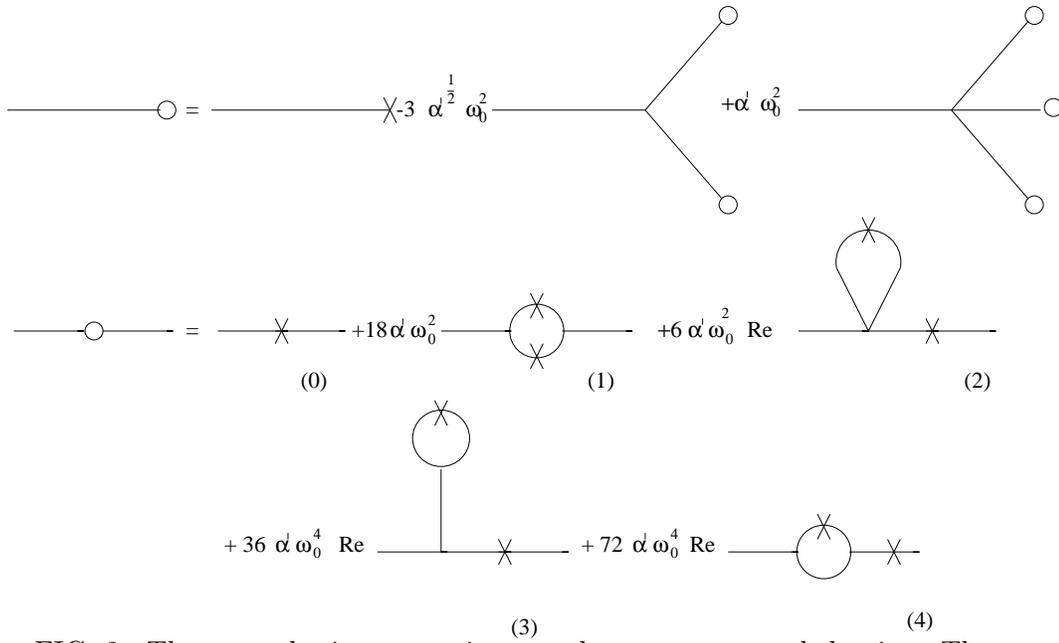,width=400pt,angle=-90}
\caption{The perturbative corrections to the power spectral density. 
The graphs are obtained using the Feynman formalism for the correlation 
function, similar to a ``propagator''}
\label{fig:ac}
\end{figure}

The Feynman diagrams are
shown in Fig.\ref{fig:ac} with the graphic representation of the solution. When
two $x'(t)$ are connected, the result is a tadpole type diagram. Indeed the mean
value of $x'^2(t)$ does not depend on $t$ if the conditions are stationary. In
this context it should be reminded that the fluctuation-dissipation mechanism
has been extended to the perturbed case. This generalization is to be intended
as a limit of the methods and it is based on the physical assumption that the
results are slightly modified as it is assumed in field theory techniques.

The final result is
\begin{eqnarray*}
<x'(f)x'(f')>&=&\delta(f+f'){2\omega_{0}^{3}\over Q}\frac{1}{(\omega^2-2
\omega_{0}^{2})^2+{\omega^2\omega_{0}^{2}\over Q^2}}\left\{1+\frac{12\alpha'
\omega_{0}^{2}(2\omega_{0}^{2}-\omega^2)}{(\omega^2-2\omega_{0}^{2})^2+
{\omega^2\omega_{0}^{2}\over Q^2}}+\right. \\ && \frac{18\alpha'\omega_{0}^{4}}
{(\omega^2+{\omega_{0}^{2}\over Q^2})[(\omega^2-8\omega_{0}^{2})^2+{4\omega^2
\omega_{0}^{2}\over Q^2}]}\left[\frac{(\omega^2+{4\omega_{0}^{2}\over Q^2})
(\omega^2+{\omega_{0}^{2}\over Q^2})-64\omega_{0}^{4}-{24\omega_{0}^{4}\over 
Q^2}}{\omega^2-8\omega_{0}^{2}+{\omega_{0}^{2}\over Q^2}}+ \right. \\ && \left.
\frac{{32\omega_{0}^{4}\over Q^2}(\omega^2-\omega_{0}^{2})+{\omega^2
\omega_{0}^{2}\over Q^2}({4\omega_{0}^{2}\over Q^2}-3\omega^2)+\omega^2(10
\omega^2\omega_{0}^{2}-16\omega_{0}^{4}-\omega^4)}{(\omega^2-2\omega_{0}^{2})^2
+{\omega^2\omega_{0}^{2}\over Q^2}}\left.\right]\right\}
\end{eqnarray*}
where the first line represents the graphs (0),(2) and (3) in Fig.\ref{fig:ac}
and the corresponding contribution modifies the definitions of the parameters 
but does not introduce a different dependence on time. This means that the 
typical shape of the power spectral density is not changed excepted some small
corrections of its parameters. 

The characteristics of the spectrum are indeed changed by the other
contributions involving convolution integrals \cite{MN}.
These terms correspond to the graphs (1) and (4) that introduce a new feature
in the shape of the spectrum. Indeed, they modify the power spectral density 
at low frequency for ${\omega_0\over Q}<<\omega<<\omega_0$. In this range the 
spectrum decreases with increasing $\omega$. The spectral density changes its
shape when the value of various parameters is modified, but there is still a 
tail for that interval falling as $\omega^{-\gamma}$ for $0<\gamma<2$. 

Moreover a new feature accounted for in the formula above is a new peak that 
is produced at a frequency which is approximately twice the value of the main
peak frequency. All these analytical results are obtained using a perturbative
method that assumes that the system is never very far from the equilibrium 
conditions. 

If the system is allowed to go back and forth between the two minima of the 
potential function, an alternative approach is needed in order to simulate 
the dynamics of the system. A simulation code has been carried out and used to
analyse different interesting cases.

The physical picture is that the impurity has some chance of oscillating
between two equilibrium positions acquiring energy from thermal excitations.

It may be pushed into one direction or the other isotropically and the 
representation of those fluctuations is simply given by the stochastic term. 

If some kicks are in the same direction, energy is gained to overcome the 
potential barrier. This model reproduces the competition between thermal 
excitation and maxima between wells in the potential function. If the 
temperature is high the passage through the barrier has more probabilities to
occur as it might be expected since $T$ is a statistical definition of energy.

At this point there is no straightforward way for keeping the damping term in
the non-linear regime, apart from making physical guesses about the type of 
damping process. Choosing other forms would mean different spectral densities
since the dynamics would be different.

The relation between the random force and the damping term that is defined by
the fluctuation-dissipation theorem in the linear approximation, has been used
in the stochastic equation outside the two regions near the minima as well. It
provides a solution whose distribution in stationary conditions satisfies the
Boltzmann one.  In general the Langevin equation implemented in the simulation
code corresponds to a Fokker-Planck equation for the probability distribution
that has the Boltzmann distribution as its stationary solution. This is one of
the motivations to keeping the same expressions for the damping force and the 
random term in the non-linear case. The other motivation is that the nature of
the damping effect is often defined by the properties of the heat-bath rather 
than on the mechanical characteristics of the physical system in contact with
it. Another point worthy of noting is that the natural scale for the x-axis is
related with the magnitude of the fluctuating forces. This is why the variable
$x'=x\sqrt{m\omega_{0}^{2}\over{K_BT}}$ has been used throughout our analyses.

\section{Simulation and Numerical Results}

The analogy existing between
the Schr\"oedinger equation and the Fokker-Planck equation can be exploited 
and an algorithm may be written based on this similarity. In particular the 
drift term and the diffusion term in the Fokker-Planck equation, play the same
role of the potential function and kinetic term in the Hamiltonian operator. 

Using the same rules for path-integral calculations, the evolution operator 
may be factored in such a way that successive steps are performed which make 
the system fluctuate according to the diffusion term or make a step according 
to the deterministic term. 

Depending upon the required precision of the stepping in the simulated states
of the system, a particular number of intermediate steps is needed to make the
system evolve from its state to the successive one. This technique is simply 
an application of the Baker-Campbell-Hausdorff formula to a stochastic 
phenomenon
\cite{BC,BD}.

\begin{figure}
\psfig{file=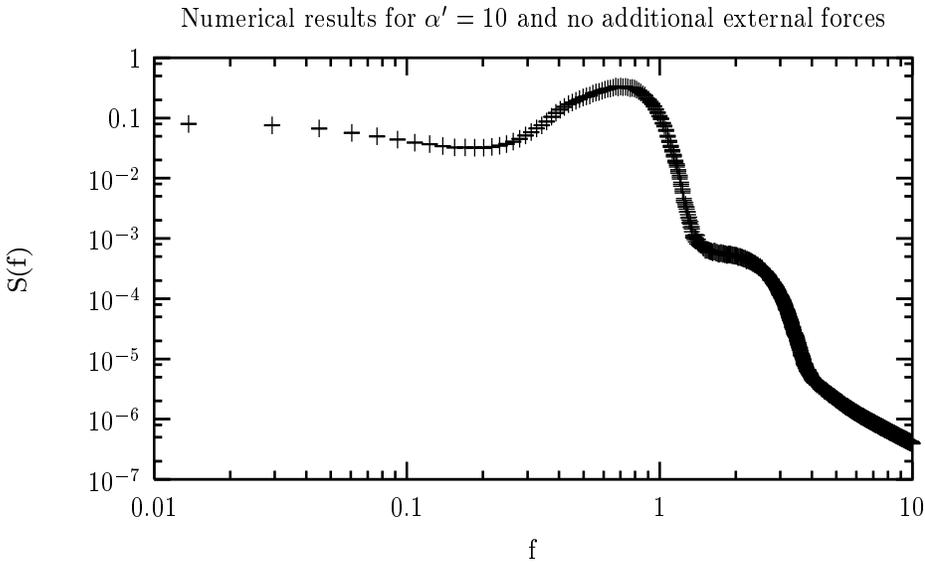,width=350pt,angle=-90}
\caption{The power spectral density is shown for $\alpha'=10$. Many values 
have been tested for $\omega_0$ and $Q$ which only influence the shape of the
peaks. For this simulation the values chosen are $\omega_0=2.7Hz$ and $Q=100$}
\label{fig:ad}
\end{figure}

Recall that large values of $\alpha'$ imply high $T$ or flat wells.
In this high energy regime, the dynamics at low frequency are such that there 
is a constant trend as in harmonic oscillators damped by a viscous force. 
In Fig.\ref{fig:ad} the spectral density is shown for $\alpha'=10$. No tail is
appreciated at low frequency.

\begin{figure}
\psfig{file=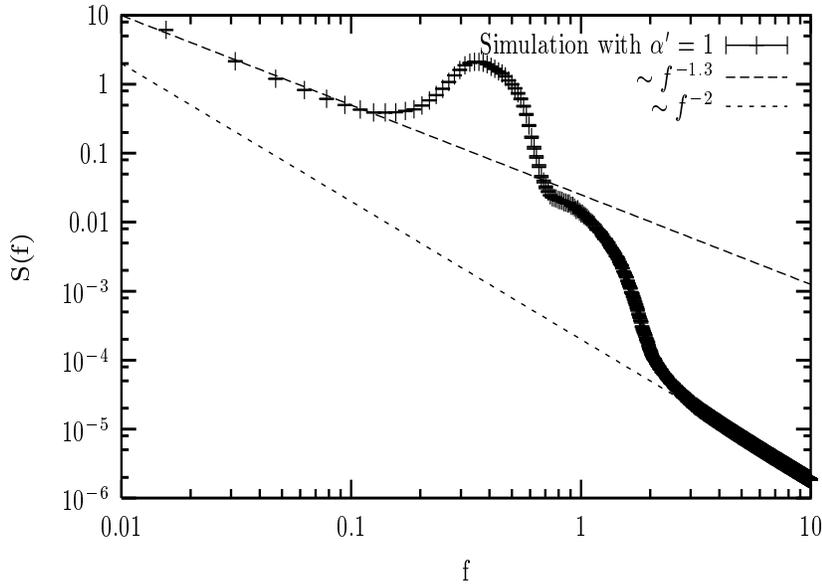,width=350pt,angle=-90}
\caption{The power spectral density is shown for $\alpha'=1$, $\omega_0=2.7Hz$
and $Q=100$ and these results can be compared to the ones in Fig.\ref{fig:ad}.
The two wells are deeper and the curve gains a tail at low frequency that is 
$f^{-1.3}$. 
After reaching stationary conditions the averages over 5000 runs were computed}
\label{fig:ae}
\end{figure}

If the value of $\alpha'$ is decreased the shape of the potential function is 
deeply characterized by two well defined minima and this affects the dynamics
at low frequency where the trend of the power spectrum $S(f)$ is not constant.
On the contrary there is a slope at low frequency that in Fig.\ref{fig:ae} is
quite visible and fitted with $f^{-1.3}$ for the specific case of $\alpha'=1$.
Indeed according to the tests I made using different values for the parameter
$\alpha'$ the low frequency tail has the form $f^{-\gamma}$ with $0<\gamma<2$.

The dynamical model we constructed shows a low frequency tail in its spectrum.
How is this system affected by a frequency independent damping force? Will the
low frequency tail be attenuated by this external force? This system is a good
candidate for investigation as the main fundamental modes are easily detected
and suppressed by actively feeding back to the resonant motion. The damping we
will use is an external viscous force proportional to $\dot x'$. Our aim is to
find out whether an external damping mechanism can suppress $S(f)$ at low $f$.

Some problems are to be faced in numerical studies. The numerical fluctuations
in simulations are the statistical errors that affect the mean value. They are
reduced by averaging over many sets of data in order to reduce the deviations

The statistical fluctuations are especially large at low frequency so that if
we are interested in obtaining a smooth curve with small error bars, many runs
must be averaged over.

\begin{figure}
\psfig{file=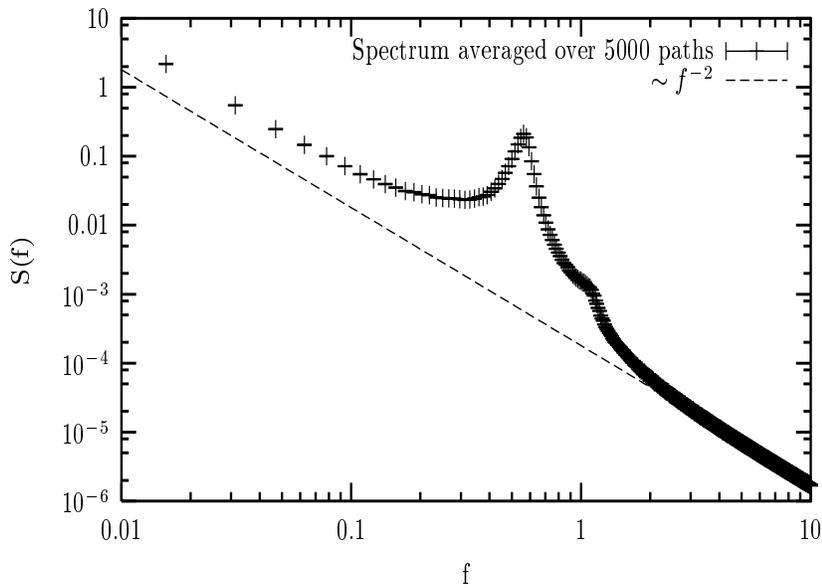,width=350pt,angle=-90}
\caption{An external damping force is introduced and its impact on the 
spectral curve is concentrated in the range of frequency around the peaks. 
Indeed one of them is substantially depressed}
\label{fig:af}
\end{figure}

In Fig.\ref{fig:af} the spectral density is shown when the damping is applied.
Only one of the peaks is well defined and indeed it becomes sharper since the
damping force limits the motion to small oscillations around one of the minima
of the potential function.

The situation at low frequency is more interesting. The values at $f\sim0$ are
not affected. In the range below the peak the low frequency tail is changed as
$f^{-2}$. A similar trend may be obtained by increasing the depth of the wells
so that the motion is mostly confined to oscillations inside one of the wells.

Many models have been proposed which involve a tail $f^{-1}$, as it is typical
of electronic circuits' noise.

For mirrors the value of $\gamma$ for a spectrum $f^{-\gamma}$ is not certain.
In fact, making experimental measurements of the fluctuations at low frequency
is difficult even for state-of-the-art optics reasearch. Experimental data are
mostly based upon resonance measurements and spectral curves are extrapolated
for very low frequencies.

For lack of off-resonance measurements linear equations have been studied with
a complex elastic force \cite{Ph,Vi,Ps}.

The random-walk problem studied in this work focuses on only one property that
is the existance of more than one equilibrium configuration. A similar model 
is used to reproduce the switching of polymeric units between two positions 
which are energetically equivalent. 

\begin{figure}
\psfig{file=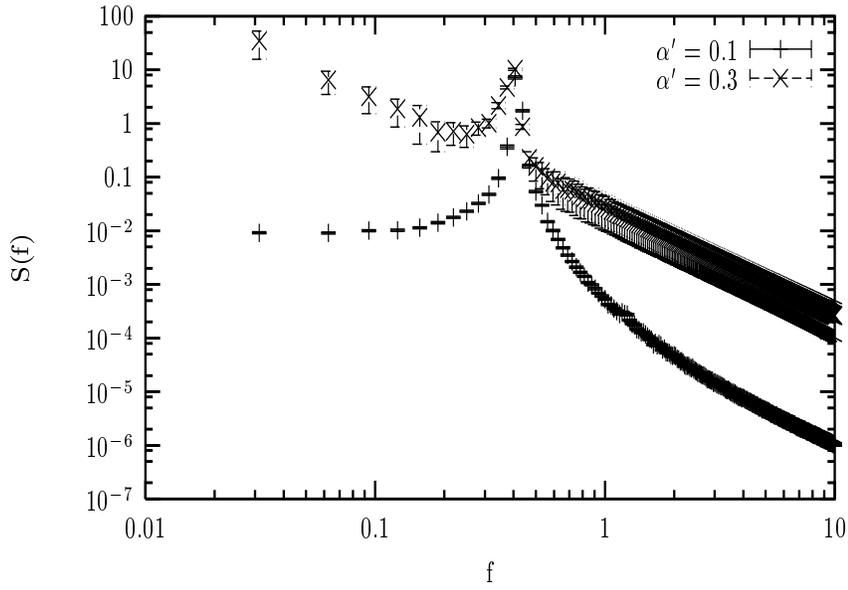,width=350pt,angle=-90}
\caption{Another example of tuning a constant to have a low frequency tail. 
The deterministic force is periodic on a lattice unidimensional space. 
The same scaling law applied previously, that takes the temperature into 
account, is used here to make quantities dimensionless}
\label{fig:ag}
\end{figure}

The simulation program has also been modified in order to investigate what may
happen if there are more than two minima for the potential function. Instead 
of the two wells, a periodic potential has been studied giving the same 
results. In Fig.\ref{fig:ag} the spectra obtained for two values of $\alpha'$ 
are shown. The fluctuation-dissipation theorem is extended to the general case
although it could only be applied in the linear approximation near one of the
minima of the potential function. Also $x$ may be scaled in order to have $x'$
in the potential function 
$$V(x')=\omega_{0}^{2}(1-\cos\sqrt{\alpha'}x')\qquad.$$

From the physical point of view a potential barrier is needed to represent the
limits of the periodic structure that otherwise would be infinitely large, and
this is not realistic.

In this case a normalized stationary solution of the Fokker-Planck equation is
the corresponding Boltzmann distribution which takes into account the modified
potential function. In other words what is needed is a differentiable term that
provides $V(x')\rightarrow\infty$ for both the limits $x'\rightarrow\pm\infty$
\cite{Pa}.
 
We assumed that the size of the periodic structure is so large that no border
condition affects the model.

\section{Conclusions}

This unidimensional model is characterized by a tail at low frequency in its 
power spectral density that is not modified even when a frequency independent
damping force is applied.

The data are obtained from a time domain simulation. We have considered both a
simple case for a two wells potential function and the more general case of a
periodic frame with oscillations between the minima. The fluctuating motion of
the degree of freedom and the associated dissipation are a model to produce a
typical low frequency tail $f^{-\gamma}$ in the power spectral density $S(f)$.

The slope of such low frequency tail can be tuned by varying the constants; it
depends on the competition between the temperature and the depth of the wells.

\section*{Acknowledgements}

Thanks are due to Director Barry C. Barish and Deputy Director Gary H. Sanders
of Ligo Laboratories for generously giving me the opportunity of going on with
my studies and Dr. Riccardo DeSalvo for mediating my approach to Ligo Project.

\end{document}